\documentclass[10pt,twocolumn]{article}

\usepackage{amsmath,amsfonts,amssymb}

\newcommand{\be}{\begin{equation}}
\newcommand{\ee}{\end{equation}}
\newcommand{\ba}{\begin{eqnarray}}
\newcommand{\ea}{\end{eqnarray}}

\newcommand{\ns}{\normalsize}

\newcommand{\ax}{\alpha}
\newcommand{\bx}{\beta}
\newcommand{\cx}{\gamma}
\newcommand{\dx}{\delta}
\newcommand{\ox}{\omega}

\newcommand{\gx}{\gamma}
\newcommand{\ab}{\bar\alpha}

\newcommand{\Ox}{\Omega}

\newcommand{\W}{\mathcal W}
\newcommand{\V}{\mathcal V}
\newcommand{\wg}{\wedge}
\newcommand{\NK}{nearly K\"ahler}

\begin{document}

%\begin{titlepage}

\setcounter{footnote}{1}
\title{\ns
   \hfill{\ns hep-th/0409008\\}
   \vskip .5cm
   {\Large \bf Heterotic compactifications and nearly-K\"ahler
   manifolds}\\[0.5cm]
  {\large Andrei Micu\footnote{On leave from IFIN-HH Bucharest.}\\[.2cm]}
  {\it\ns Department of Physics and Astronomy, University of
   Sussex}\\
  {\it\ns Brighton, BN1 9QH, United Kingdom.}\\
   {\it\ns email: A.Micu@sussex.ac.uk} \\
\begin{quote}
{\small \quad We propose that under certain conditions heterotic string
   compactifications on half-flat and \NK{} manifolds are equivalent.
   Based on this correspondence we argue that the moduli space of the
   \NK{} manifolds under discussion consists only of the K\"ahler
   deformations moduli space and there is no correspondent for the
   complex structure deformations.}
\end{quote}
\vspace{-1.5cm}
}
\date{}

\maketitle

\noindent
Recently, a lot of attention was devoted to the study of
compactifications on manifolds with reduced structure group. Such
manifolds usually appear when one considers generalized
compactifications in the presence of non-trivial fluxes. It is
well-known that purely geometrical compactifications are
supersymmetric if the internal manifold admits a covariantly constant
spinor or in other words it has special holonomy \cite{DPN}. When
background fluxes are present these 
solutions are deformed away from the special holonomy limit and the
internal manifold only has reduced structure group
\cite{GMSW1}--\cite{BC2}. 
A second way to see these manifolds appearing is via dualities. In
particular applying some duality transformation to a background which
has non-trivial NS-NS fluxes generically leads to a deformed geometry
\cite{GLMW}--\cite{ABBDKT}. The bottom line of both
approaches is that in 
such compactifications a superpotential is generated and some of the
geometric moduli get stabilized
\cite{BBDG,BBDP,CCDL1,GLMW,GM,ABBDKT,GLM}. In a very recent work,
\cite{GLM}, it was shown by an explicit calculation that for the case
of heterotic strings compactified on manifolds with $SU(3)$ structure
the superpotential has a very simple form\footnote{See also
  Refs.~\cite{BBDP,CCDL1,ABBDKT}.} 
\begin{equation}
  \label{WGLM}
  W = \sqrt 8 \int \Ox \wg (H + i dJ) \; ,
\end{equation}
where $\Ox$ and $J$ are the $SU(3)$ invariant forms on a manifold with
$SU(3)$ structure and $ H$ is the field strength of the two-form
field $ B$. What is striking about this superpotential is that
even if no particular assumption was done to obtain this expression,
i.e. it is valid for \emph{any} manifold with $SU(3)$ structure, the
superpotential seems to depend explicitly only on the first torsion
class $\W_1$ of the manifold under consideration (see
Ref.~\cite{CS} for a description of the five torsion classes of
manifolds with $SU(3)$ structure).

In this note we plan to elaborate more on this issue, but do not keep
the discussion general and instead concentrate on the half-flat
manifolds which were considered in Ref.~\cite{GLM}. Note first that the
K\"ahler potential does not depend on the torsion of the internal
manifold and is the same as in the torsionless Calabi--Yau
compactifications. Thus, if the above observation that the
superpotential only depends on the torsion class $\W_1$ is true, it
implies that the whole low energy physics depends just on this first
torsion class and all the other torsion components are completely
irrelevant from a four dimensional point of view. This is a curious
statement, but also quite interesting as it allows us to find
manifolds which are simpler than the ones considered in Ref.~\cite{GLM} and
which produce the same result in four dimensions. The only requirement
seems to be that the first torsion class, $\W_1$, of these new
manifolds is the same as the first torsion class of the manifolds
considered in Ref.~\cite{GLM}.

Among the manifolds with $SU(3)$ structure the simplest ones are the
so called \NK{} manifolds (also known as manifolds with weak $SU(3)$
holonomy) for which only the torsion component in
$\W_1$ is non-vanishing. On such manifolds the $SU(3)$ invariant forms
$J$ and $\Ox$ satisfy
\begin{equation}
  \label{dJO}
  dJ \sim \Ox \; ,
\end{equation}
and the value of the proportionality constant in this equation is
given precisely by the torsion. Such manifolds have appeared recently
in string compactification in Refs.~\cite{BC1,LS,BC2}
and it would be interesting to learn how to handle theories
compactified on such manifolds. The first reason is that they are much
simpler than the half-flat manifolds previously considered in the
literature and moreover one has a better control on their geometry as
the Ricci tensor and scalar can be computed explicitly. It is also
worth noting that the half-flat manifolds discussed in Ref.~\cite{GLMW} can
not be reduced to \NK{} manifolds as in this case the torsion classes
are related and one can not set them independently to zero.
Thus, the question we will be focusing in this note is:
\begin{quote}
  Does there exist a nearly K\"ahler manifold such that the
  compactification of the heterotic string at the lowest order in
  $\ax'$ yields the same result as the one obtained in Ref.~\cite{GLM}?
\end{quote}
To answer this question we would have to derive the low energy action
of the heterotic string\footnote{In this note we do not consider at
  all the gauge fields and so the gauge group is irrelevant for the
  discussion.} compactified on a \NK{} manifold and compare it with
what was obtained in Ref.~\cite{GLM}.

%%%%%%%%%%%%%%%%%%%%%%%%%%%%%%%%%%%%%%%%%%%%%%%%%%%%%%%%%%%%%%%%%%%%%%
%%%%%%%%%%%%%%%%%%%%%%%%%%%%%%%%%%%%%%%%%%%%%%%%%%%%%%%%%%%%%%%%%%%%%%

Before we start there is a subtlety which we should clarify.  The
assertion that the superpotential \eqref{WGLM} only depends on $\W_1$
is not quite true. To see this we integrate Eq.~\eqref{WGLM} by parts to
obtain
\begin{equation}
  W = \sqrt 8 \int d\Ox \wg \big(B + i J \big) \; .
\end{equation}
By definition the term $\int d \Ox \wg J$ is the component of the
torsion in the first torsion class \cite{CS}, but $\int d \Ox \wg B$
can in general depend on the other torsion classes as well. Indeed,
there is no reason for which $B$ should be a singlet under $SU(3)$.
There is however one exception to this, namely when there is only one
K\"ahler modulus i.e. when the underlying a Calabi--Yau manifold has
$h^{(1,1)} = 1$. Recall that the prescription for obtaining the
moduli fields on a half-flat manifold is to expand the forms $B$
and $J$ in a basis of $(1,1)$ forms. Thus we write
\begin{equation}
  \label{mod}
  \begin{aligned}
    J = & \; v \, \ox \; ,\\
    B = & \; b  \, \ox \; ,
  \end{aligned}
\end{equation}
and we see that $B$ and $J$ are actually proportional to each
other. Consequently $B$ is also a singlet under the structure group
and so the superpotential \eqref{WGLM} only depends on the torsion
class $\W_1$ as we expected. 

Restricting to this special case is not such a bad choice. First of
all we know that the are examples of Calabi--Yau manifolds with
$h^{(1,1)}=1$, \cite{GSW}, and thus we can argue using the
construction presented in Ref.~\cite{GLMW} that the corresponding
half-flat manifolds with the properties discussed above also
exist. Moreover, many of the 
aspects of the heterotic strings compactified on Calabi--Yau manifolds
are much easier handled for this special case of $h^{(1,1)}=1$ and so
we expect that the analysis in this note will not be completely
irrelevant for this subject.

In this limit, the $N=1$ supergravity obtained in Ref.~\cite{GLM} can be
described in terms of the following K\"ahler potential 
\begin{equation}
  \label{Kpot}
  K = K_T + K_S + K^{(cs)} \; ,
\end{equation}
where
\begin{eqnarray}
  \label{KT}
  K_T & = & -3 \ln{(i(\bar T - T))} \; , \\
  \label{KS} 
  K_S & = & - \ln{(i(\bar S - S))} \; , \\
  \label{Kcs}
  K^{(cs)} & = &- \ln{i \int \Ox \wedge \bar \Ox} \equiv - \ln{\left(\V
  ||\Ox||^2 \right)} \; .
\end{eqnarray}
The superpotential has the form
\begin{equation}
  \label{W1}
  W = \sqrt 8 e T \; ,
\end{equation}
and the complexified K\"ahler modulus is given as usual by $T= b +
iv$. As argued before, this result only depends on the first torsion
class $\W_1$ of the half-flat manifold\footnote{Note that in this
  case, the fact that there are no other $(1,1)$ forms except $\ox$
  which appears in \eqref{mod}, the torsion in the second class $\W_2$
  also vanishes.} and we would like to see what is the precise role of
the other torsion classes on which this low energy action does not seem
to depend. The way we propose to answer this question is by studying
the compactification of the heterotic strings on a nearly K\"ahler
manifold which would reproduce the above K\"ahler potential and
superpotential. One of the major problem we face in this approach is
the fact that the moduli space of manifolds with $SU(3)$ structure in
general, and \NK{} manifolds in particular, is not known. On the other
hand in the simpler case we are considering, namely that there is one
single K\"ahler modulus,\footnote{% 
  Of course one can ask the question if such \NK{} manifolds do indeed
  exist. We are not aware of a precise answer to this question, but we
  know that one of the simplest examples of a \NK{} manifold is $S^6$
  which definitely has this property. However, the sphere does not
  have only one spinor which is globally defined and so this example
  does not really fall into the class of manifolds we discuss in this
  paper.}
the expansions \eqref{mod} are almost the only things one can write down.
As a second argument for keeping the expansions \eqref{mod} is that we
would again like to think of these manifolds as small deformations
of some underlying Calabi--Yau manifold on which one generically
writes the moduli expansions \eqref{mod}. 
%% Last, but not least, Using a
%% similar argument to \cite{GLMW}, the fields $b$ and $v$ in \eqref{mod}
%% will have masses of order $e$ which we assume is much smaller than the
%% Kaluza-Klein scale, and thus we want to keep in our analysis.

It is important to note that once we make this choice, the K\"ahler
potential and the superpotential are given by the general analysis
presented in Ref.~\cite{GLM} and in particular they will have precisely the
form \eqref{Kpot} and \eqref{WGLM}. The only thing we have to fix is
the torsion of this \NK{} manifold which should be chosen in such a
way that the superpotential actually reproduces Eq.~\eqref{W1}. As
explained before, specifying the torsion for such manifolds amounts to
make the relation \eqref{dJO} more precise. Since $dJ$ is a real form,
let us try\footnote{The choice of imaginary or real part of the form
  $\Ox$ is pure convention.}
\begin{equation}
  \label{dJ}
  dJ = \frac{2 (ev) \Ox_-}{i \int \Ox \wg \bar \Ox} \; ,
\end{equation}
where $\Ox_-$ is the imaginary part of the form $\Ox$ and is defined
as  
\begin{equation}
  \label{O-}
  \Ox_- = \frac{\Ox - \bar \Ox}{2i} \; .
\end{equation}
Using this, and equation \eqref{mod} we see that the basis form $\ox$
satisfies a similar relation
\begin{equation}
  \label{dox}
  d \ox = \frac{2 e \Ox_-}{i \int \Ox \wg \bar \Ox} \; .
\end{equation}
Inserting Eq.~\eqref{mod} in Eq.~\eqref{WGLM} and using
Eq.~\eqref{dox} one can easily see that the superpotential is indeed
given by \eqref{W1}.

To check that this set-up is correct we would have to compute the
potential and see if it agrees with the one found in 
Ref.~\cite{GLM}.\footnote{Here
we implicitly assume that the kinetic terms of all the fields are the
same as in the half-flat compactification presented in Ref.~\cite{GLM} and
thus possible differences can come only from the potential.}
For the computation which follows we use the conventions and the
ten-dimensional action for the heterotic strings given in
Ref.~\cite{GLM}. 

Again, like in Ref.~\cite{GLM}, there will be two contributions to the
potential: one coming from the kinetic term of the NS-NS field $B$ in
ten dimensions and the second from the curvature of the internal 
manifold. Let us compute them separately.

Using Eq.~\eqref{dox} the internal part of $H$ is given by
\begin{equation}
  H_{int} = 2 (be) \frac{\Ox_-}{i \int \Ox \wg \bar \Ox} \; .
\end{equation}
By a standard calculation one immediately finds
\begin{equation}
  \label{VH}
  \begin{aligned}
  V_H \; & = \frac{e^{2 \phi}}{4 \V} \int H_{int} \wg * H_{int}  \\
  & = 4 e^{2 \phi + K_T + K^{cs}} (eb)^2 \; ,    
  \end{aligned}
\end{equation}
where we used Eq.~\eqref{Kcs} and \eqref{O-}, and the fact that
\begin{equation}
  8 \V = i (T-\bar T)^3 = e^{K_T} \; .
\end{equation}

The computation of the scalar curvature for \NK{} manifolds is not
difficult either. In fact one can compute the full Ricci tensor as
these manifolds are known to be Einstein \cite{AN}. 

There are probably many ways to compute the curvatures of a \NK{}
manifold, but here we will follow Ref.~\cite{BDS} where the Ricci
tensor for manifolds with weak $G_2$ holonomy was computed by making
use of the Killing spinor equation. To find this, note first
that the (con)torsion of \NK{} manifolds is a totally antisymmetric
tensor and using Eq.~\eqref{dJ} one finds
\begin{equation}
  \label{ct}
  \kappa_{mnp} = \frac13 e^{K^{(cs)}} (\Ox_+)_{mnp} \; .
\end{equation}
Thus, the relation the $SU(3)$ invariant spinor\footnote{We use the
  convention that $\eta$ is a Weyl spinor of positive chirality.} $\eta$
satisfies, is 
\begin{equation}
  \label{Deta}
  \nabla_m \eta = \frac{e^{K^{(cs)}}}{12} (\Ox_+)_{mnp} \gx^{np} \eta
  \; .
\end{equation}
In order to compute the Ricci scalar using this relation it will be
useful to expand the right hand side in the standard basis for spinors
defined as
\begin{equation}
  \eta \; , \quad \gx_{\ab} \eta \; , \quad \gx_{\ax} \bar \eta \; , 
  \quad \bar \eta \; ,
\end{equation}
where $\ax$ denote complex indices which run $\ax = 1 , \ldots , 3$.
Noting that the right hand side of Eq.~\eqref{Deta} has positive
chirality, the most general expansion we can write is
\begin{equation}
  \label{Detaexp}
  \begin{aligned}
    \nabla_{\ax} \eta \; & = \frac{e^{K^{(cs)}}}{12} (\Ox_+)_{\ax \bx \cx}
    \gx^{\bx \cx} \eta \\ 
    & = A_{\ax} \eta + A_\ax^\bx \gx_\bx \bar \eta \; .
  \end{aligned}
\end{equation}
Multiplying successively by $\bar \eta^T$ and $\eta^T \gx^\dx$ one obtains
\begin{equation}
  A_{\ax} = 0 \; \, \qquad A_\ax^\bx = i \frac{e^{K^{(cs)}} \sqrt 2}{12}
  ||\Ox|| \dx_\ax^\bx \; ,
\end{equation}
and thus, Eq.~\eqref{Deta} becomes
\begin{equation}
  \nabla_m \eta = i \frac{e^{K^{(cs)}} \sqrt 2}{12} ||\Ox|| \gx_m \eta
  \; .
\end{equation}
From here on the calculation for the scalar curvature is
straightforward. Taking the commutator of two covariant derivatves
acting on the spinor $\eta$ and using standard gamma-matrix
relations the result is found to be\cite{BDS}\footnote{Note that the
  different sign comes from the different convention we use for the
  Ricci scalar.}
\begin{equation}
  \label{R}
  R= - \frac53 e^{2K^{(cs)}} (ev)^2 ||\Ox||^2 \; .
\end{equation}
After integrating over the internal manifold one finds the
contribution to the potential which comes from the curvature to be
\begin{equation}
  \label{VR}
  \begin{aligned}
  V_R \; & = \frac{e^{2\phi}}{2 \V} \int \sqrt g R \\
  & = - \frac{20}{3} e^{2 \phi + K^K + K^{(cs)}} (ev)^2 \; .    
  \end{aligned}
\end{equation}

Putting together the results from Eq.~\eqref{VH} and \eqref{VR} we obtain
\begin{equation}
  \label{V}
  V = 4 e^{2 \phi + K^K + K^{(cs)}} \Big[(eb)^2 - \frac53(ev)^2 \Big] \; .
\end{equation}
Let us now compare this potential with the one derived in
Ref.~\cite{GLM}. For the case $h^{(1,1)}=1$ the inverse metric on the
K\"ahler moduli space can be computed from the K\"ahler potential
\eqref{KT} and is given by $\frac43 v^2$. With this, the formula for
the potential in Ref.~\cite{GLM} becomes 
\begin{equation}
  \label{Vhf}
  V_{hf} = 4 e^{2 \phi + K^K + K^{(cs)}} \Big[4 (eb)^2 + \frac43(ev)^2
  \Big] \; .
\end{equation}
Clearly the two formulae are different, but what had gone wrong
because as we have argued at the beginning the low energy actions
should have been the same as long as the first torsion classes were
chosen to be the same. Thus we probably missed something in the above
argument. 

One of the things we have not discussed at all is the moduli spaces of
half-flat and \NK{} manifolds. On the half-flat manifolds used in
Ref.~\cite{GLM} one assumes that the moduli space of metrics is
identical to the moduli space of some underlying Calabi--Yau manifold.
Can this still be assumed for \NK{} manifolds as we implicitly did?
The answer to this question is quite difficult as now we do not have
any guiding principle in order to make such assumptions as it was the
case for half-flat manifolds where mirror symmetry was enforcing the
structure of the moduli spaces \cite{GLMW}. There is however one thing
we can speculate on in the case of \NK{} manifolds.  Equation
\eqref{dJ} tells us that $\Ox_-$ is an exact form and thus its
corresponding cohomology class is trivial.

For the moduli space of complex structures we were assuming that it
can be again described in terms of the form $\Ox$.  However, if this
is bound to be in the trivial class by the relation \eqref{dJ} it
means that its variation with respect to the complex structure moduli
can only be exact or in other words that \NK{} manifolds do not
allow deformations which change $\Ox$ and thus they do not have what
we called by abuse of language a complex structure moduli space.

This is quite intriguing so let us see if it can be independently
verified by the low energy analysis we have presented so far. For this
we would further have to specialise to a `rigid' half-flat manifold
which does not have any complex structure deformations. In
Ref.~\cite{GLM} it was argued that the contribution to the potential
from the complex structure moduli is $3 |W|^2$. It is easy to see that
subtracting this contribution from the potential \eqref{Vhf} one
indeed recovers the potential \eqref{V}.

To summarize we have learned that the low energy action for the
$(\ax')^0$ sector of the heterotic string on a \NK{} manifold has the
form 
\begin{eqnarray}
  \label{Sfin}
  S_{nK} & = & \; \int \Big\{ -\frac12 R *1 - d \phi \wg * d \phi - \frac14
  d a \wg * d a \nonumber \\    
  & &- \frac{3}{4 v^2} d T \wg * d \bar T -V *1 \Big\} \; ,    
\end{eqnarray}
with the potential $V$ given in \eqref{V}.
Moreover the above comparison to the heterotic theory compactified on
a half-flat manifold shows that this action is the bosonic part of an
$N=1$ supergravity coupled to two chiral multiplets $S$ and $T$ with
K\"ahler potential
\begin{equation}
  K = - \ln{(i ( \bar S - S))} - 3 \ln{(i ( \bar T - T))} \; ,
\end{equation}
and superpotential 
\begin{equation}
  W = \sqrt 8 e T \; .
\end{equation}

\vspace{.2cm}

Let us conclude this note by analyzing the main points derived so far.
One of the most surprising results is that the the (almost)complex
structure of \NK{} manifolds is rigid and that there are no moduli
associated to it. This is an interesting result in its own as it means
that on such manifolds one only has to find a mechanism to fix the
K\"ahler moduli. Moreover, proving this claim in general using a
rigorous mathematical aproach would be an important step forward in
understanding the role of manifolds with $SU(3)$ structure in string
compactifications.  However, one has to keep in mind that in obtaining
this result we assumed a couple of things which may restrict the
validity of the argument. The action obtained in Eq.~\eqref{Sfin}
corresponds to the compactification of the heterotic string on a
half-flat manifold which has $h^{(2,1)} = 0$. The first problem comes
from the fact that the half-flat manifolds were supposed to be the
mirror duals of some Calabi--Yau manifold with NS-NS fluxes turned on
and for this reason we need that $h^{(2,1)} > 0$. Ignoring this we
still have to bare in mind that we were working in the limit
$h^{(1,1)}=1$ and it is not certain that a Calabi--Yau manifold with
$h^{(1,1)}=1$ and $h^{(2,1)}=0$ does indeed exist. Furthermore, we were
relying on the fact that the moduli space of \NK{} manifolds has a
similar description to the moduli space of Calabi--Yau manifolds and
splits into K\"ahler and complex structure deformations which in turn
are governed by the $SU(3)$ singlet forms $J$ and $\Ox$ respectively.
Thus one of the minimal checkes for the validity of the picture
presented in this note would be to verify this claim in an independent
mathematical way. On the other hand, despite these potential problems
we believe that this is an important step forward in understanding
more complicated (and realistic) compactifications on manifolds with
$SU(3)$ structure and it would be interesting to extend the approach
in this note to less constrained examples.

There are also a couple of interesting features which we should stress
here. First it is definitely easier to deal with \NK{} manifolds and
as we have seen one can compute in a transparent way their Ricci
scalar. It is also worth noting that the expression obtained in
\eqref{R} is negative and thus one can think of configurations where
ordinary fluxes are turned on which contribute a positive energy
density and can balance the negative term coming from the curvature to
give a Minkowski vacuum in four dimensions. Note that this possibility
was not known before as the generic solution for compactifications on
half-flat manifolds are domain walls \cite{MM} and in order to obtain
a Minkowski vacuum additional constructions would be needed. However,
it is important to stress here that even if such a ground state is found
this should necesarily be non-supersymmetric as the conditions for
supersymmetry derived in Ref.~\cite{AS1} require the internal manifold
to be complex (see also Ref.~\cite{CCDLMZ} for a more recent discussion in
terms of torsion classes).  Last, but not least, it is worth noting
that such manifolds can be an interesting arena for the study of an
effective action in the presence of both electric and magnetic fluxes
as this problem has not yet been solved.  In Ref.~\cite{BDKT} it was proved
that the magnetic fluxes should come from non-vanishing derivatives of
the form $\Ox_-$. In the language of this paper this would correspond
to a \NK{} manifold for which $dJ$ in Eq.~\eqref{dJ} has also piece
proportional to $\Ox_+$. However we do not discuss this here anymore
and we leave it for a further work.

\vspace{.3cm}
\noindent
{\bf Acknowledgments} This work was supported by PPARC. I would like
thank Andre' Lukas for useful discussions and encouragement to
finish this work and for reading the manuscript. I also thank
Gianguido Dall'Agata, Klaus Behrndt and Peng Gao for comments and
useful correspondence.

\end{document}